\documentclass[]{PoS}

\title{Spectral evolution of bright NS LMXBs}

\ShortTitle{Spectral evolution of bright NS LMXBs}

\author{A. Paizis,$^a$
R. Farinelli,$^b$
L.I. Mainardi$^a$
and L. Titarchuk$^{bc}$\\
\llap{$^a$}INAF-IASF Milano, Italy\\
\llap{$^b$}University of Ferrara, Italy\\
\llap{$^c$}NASA/GSFC Greenbelt, USA\\
E-mail:\email{ada@iasf-milano.inaf.it}}

\abstract{
Theoretical and observational support suggests that the spectral evolution of neutron-star LMXBs, 
including transient hard X-ray tails, may be explained by the interplay between thermal and bulk 
motion Comptonization. 
In this framework, we developed a new model for the X-ray spectral fitting {\sc XSPEC} package which takes into account
 the effects of both thermal and dynamical (i.e. bulk) Comptonization, {\sc CompTB}.

Using data from the \textit{INTEGRAL} satellite, we tested our model on broad band  spectra 
of a sample of  persistently low magnetic field bright neutron star Low Mass \mbox{X-ray} Binaries, 
covering  different  spectral states. The case of the bright source GX~5--1 is presented here.
Particular attention is given to the transient powerlaw-like hard  \mbox{X-ray} (above 30\,keV) tail
that we interpret in the framework of the bulk motion Comptonization process, 
qualitatively describing the physical conditions of the environment in the innermost 
part of the system.}

\FullConference{The Extreme sky: Sampling the Universe above 10 keV\\
                 October 13-17 2009\\
                 Otranto (Lecce) Italy}

\begin{document}
\def\xspec{{\sc xspec}}
\def\comptb{{\sc CompTB}}
\def\comptt{{\sc CompTT}}

\section{Introduction}

\begin{figure}[h]
\centering
\includegraphics[angle=270,width=0.7\linewidth]{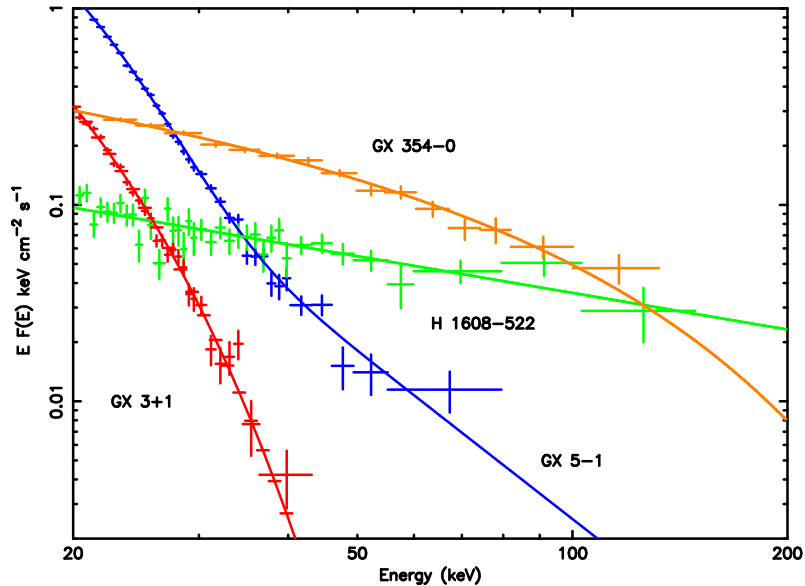}
%\vspace{4cm}
\caption{\textit{INTEGRAL} IBIS/ISGRI average spectra of GX~354--0 ("pure" low/hard state, see text), 
H~1608--522 (low/hard state), 
GX~5--1 (intermediate state) and GX~3$+$1 (very soft state) [1].
\label{fig:comp}}
\end{figure}

\begin{figure}
\centering
\includegraphics[width=0.5\linewidth]{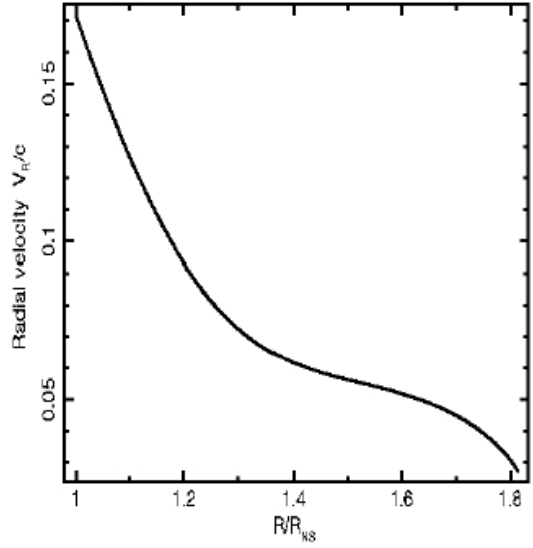}
%\vspace{4cm}
\caption{Radial velocity profile of the in-falling matter, obtained as a solution 
of the radial momentum equation for a given set of parameters which characterize 
the flow (Titarchuk \& Farinelli, in prep.)
\label{fig:profile}}
\end{figure}

Low Mass X-ray Binaries (LMXB) hosting a Neutron Star, be they transient or persistent, 
show a variety of spectral states. A collection can be seen in Figure~\ref{fig:comp}.
The "intermediate state" (spectrum of GX~5--1) is particularly interesting because it merges two different 
spectral shapes: the typical thermal Comptonization bump, characterized by a 3--5\,keV 
electron plasma temperature and optical depth ~10, and a hard tail reaching much higher 
energies than expected from a 3\,keV Comptonizing plasma. 
Such hard-tails in the spectra of NS LMXBs (so-called Z sources) have been discovered by 
\emph{BeppoSAX} 
([2] and references therein) and subsequently confirmed and discovered in 
other sources with e.g. \emph{RXTE}, \emph{INTEGRAL}. 
Shape-wise, they have been fit with a power-law while physics-wise, several models have been 
proposed: e.g hybrid corona Comptonization [3], Synchrotron emission from jet 
electrons [4], Bulk motion Comptonization (e.g. [5], [6], [7]).  In the latter model photons  
gain energy through first-order Fermi acceleration, the energization process 
producing the hard-tail occurs at the expense of the bulk kinetic energy of  the in-falling
 matter and is not supplied by the gas internal energy. 
The strong point in favour of bulk Comptonization model is that it requires no ad-hoc 
assumptions on the underlying physics (i.e. injection of non-thermal electron population by 
some unspecified mechanism)
but in fact it can be explained in terms of first 
principles, using a full magneto-hydrodynamical
treatment of the problem with appropriate
boundary conditions, in particular by solving 
the radial momentum equation (Titarchuk \& 
Farinelli, in prep.). Moreover, this solution of the 
problem in a bounded configuration can also 
explain the temporal properties of NS LMXBs 
(e.g., [8]) which are 
difficult to explain with other models. 
Figure~\ref{fig:profile} shows an example of the velocity profile 
of the in-falling matter, with a clear increase 
closer to the NS surface where bulk motion 
Comptonization becomes more important. 
A complete set of computations for a grid of main 
physical  parameters (accretion rate, etc) is in progress.

\section{A unified physical scenario}

The first self-consistent physical scenario to explain the 
spectral evolution of persistent NS LMXBs, with particular attention to the transient 
hard-tails, has been proposed by [1]: the spectral evolution was explained as the result of the interplay of two 
components, thermal and bulk Comptonization, the latter, expected in the vicinity of the NS, 
being at the origin of the hard X-ray tail. The relative contribution of the two 
Comptonization regimes is proposed to be driven by the local accretion rate. 

Thermal Comptonization is expected to originate in the external part of the corona 
(transition layer, TL) with seed photons coming mainly from the disk, while bulk 
Comptonization is produced in the inner part of the system, with seed photons from 
the NS and from the TL itself (see Figure 3).
This scenario is consistent with results 
by [8] who used 
the characteristic (break and QPO) 
frequencies of the power density 
spectrum of 4U~1728--34 to determine 
the geometric size of the configuration 
where the hard tail is formed.
 \begin{figure}
\centering
\includegraphics[width=0.9\linewidth]{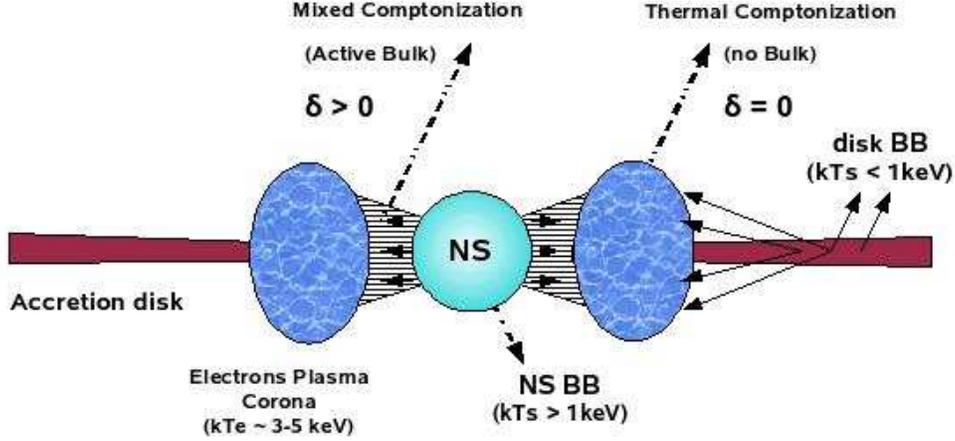}
\caption{Schematic view of the proposed scenario for thermal and bulk Comptonization 
regions in LMXBs hosting NS [9].
\label{fig:TL}}
\end{figure}

\section{A new XSPEC Comptonization model: \comptb}
We have developed a new \xspec\ model that takes into account the effects of
both thermal and bulk Comptonization [7]\footnote{\comptb, http://heasarc.nasa.gov/docs/xanadu/xspec/models/comptb.html}. 
The model consists of the sum of direct BB emission and its 
Comptonization (BB*G): 

\begin{equation}
%F(E)=\frac{C_n}{A+1}(BB + A \times BB \ast G)\label{compTB}
F(E)=\frac{C_n}{A+1}(\underbrace{BB}_{(a)} + \underbrace{A \times BB \ast G}_{(b)})\label{compTB}
\end{equation}

where $C_n$ is the normalization constant, $A$ is the illumination factor, $C_n\,BB/(A+1)$ 
is the seed photon spectrum directly seen by the observer and not modified by Comptonization processes.
 $C_n\,A/(A+1)\times BB \ast G$ is the Comptonized spectrum obtained by convolution of a seed photon spectrum 
(blackbody, BB) with the Green's Function\footnote{The Green's Function, $G$, is the response of the system to the injection of a 
monochromatic line, see [7] for details.}, in order to evaluate the effect of thermal and bulk Comptonization 
on the seed photon field.

Parameters of the model are the seed photon temperature $kT_s$, 
the electron plasma temperature $kT_e$, the spectrum energy slope (i.e. overall Comptonization efficiency) 
$\alpha$ (photon index $\Gamma=\alpha+1$), the bulk parameter 
$\delta$ that quantifies the efficiency of bulk  over  thermal Comptonization and $\log A$, which assigns a different 
weight to the two components $(a)$ and $(b)$. This model enables the co-existence of the direct seed photon
component and its Comptonized part, all obtained in a self-consistent way. In Eq.[\ref{compTB}] we note that 
for $\log A = -8$, we have only $(a)$, i.e. the direct seed photon component, while for $\log A = 8$, 
the direct component is no longer
visible and we have only thermal plus bulk Comptonization. In the case of  
$\delta=0$, the bulk Comptonization contribution is neglected, retaining
only the thermal effects (equivalent to \comptt, [10]).

\section{An application to the spectral evolution of NS LMXBs}
We have successfully applied the \comptb\ model
to several sources such as Sco~X--1, GX~340$+$0, GX~354--0 [7], GX~5--1, 
GX~349$+$2, GX~13$+$1, GX~3$+$1, GX~9$+$1 [9], GS~1826--238 [11], Cyg~X--2 [12], 
and GX~17$+$2 [13].

\begin{figure}
\centering
\includegraphics[width=1.0\linewidth]{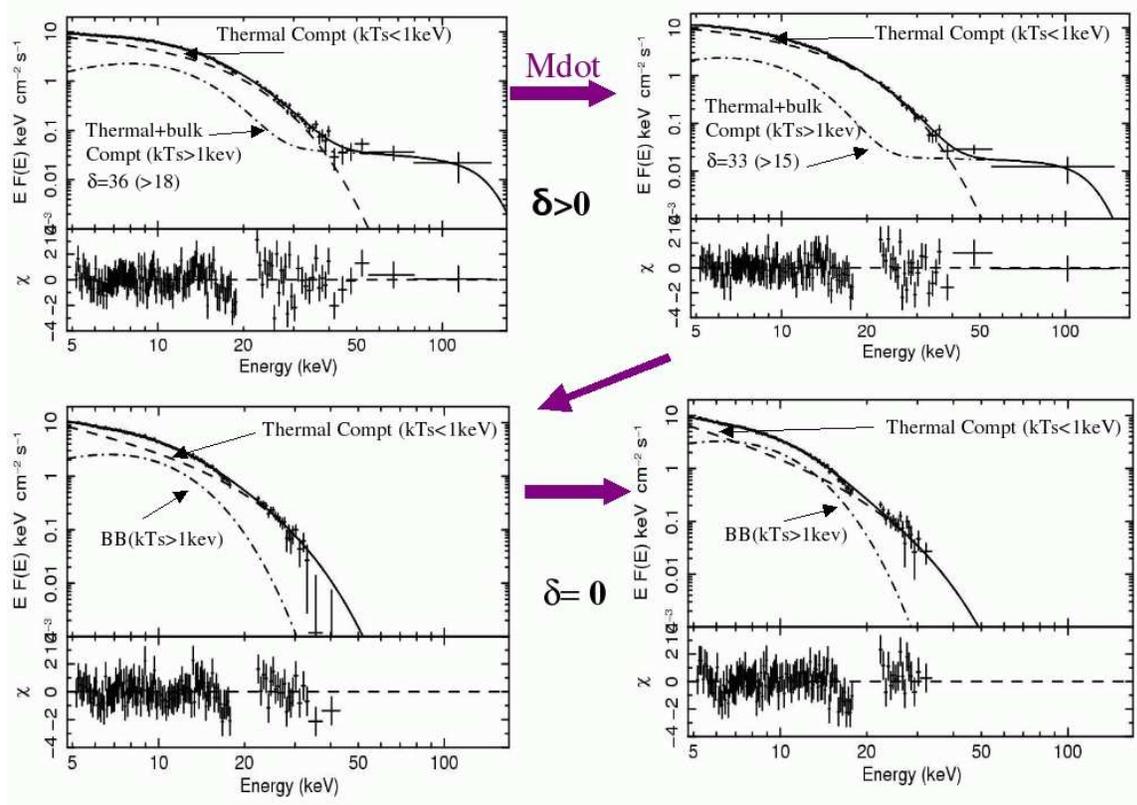}
%\vspace{4cm}
\caption{\emph{INTEGRAL} spectra of GX~5--1 fit with two \comptb\ models, one thermal, accounting 
for most of the energetic budget, and one bulk+thermal (accounting for the hard tail) 
or simple BB  [adapted from 9].
\label{fig:evol}}
\end{figure}

Figure~\ref{fig:evol} [9] shows the fit of the \textit{INTEGRAL} spectra of GX~5--1 using two
\comptb\ components: the first one (dash) is thermal Comptonization of cold disk photons 
by the outer transition layer (see Fig.~\ref{fig:TL}); the second one (dash-dot) describes the 
overall (thermal plus bulk) Comptonization of hotter seed photons close to the NS 
surface by the inner transition layer ($delta>0$, see Fig.~\ref{fig:TL}).
At increasing local accretion rates, the increasing radiation pressure from the NS inhibits the
bulk inflow stopping the Comptonization, hence the hard tail component disappears and 
we see directly the NS emission (BB with kTs $>$ 1\,keV).

\section{Conclusions}
Our approach to explain in a self consistent way NS LMXBs is giving promising results.
Understanding the origin of the hard tails in NS LMXBs is important to make a decisive step 
forward, from phenomenology (power-laws) to physics. 
This step forward will not make us understand hard tails alone, on the contrary: the hard-tails
could be the observational feature that gives us the means to understand the mechanism of 
accretion flows in general.
A large collaboration to investigate observational (e.g.  
[1], [7], [9], [11], [12], [13] and theoretical aspects (e.g. [5], [6], Titarchuk and Farinelli
in prep.) of accretion flows in X-ray Binaries is in place. 

Bulk Comptonization is gaining strong theoretical and observational support not 
only in the study of LMXB spectral evolution but also for High Mass X-ray Binaries 
(HMXB). A new theoretical model for accretion powered X-ray pulsars, based on bulk 
and thermal Comptonization occurring in the accreting shocked gas, was recently
presented by [14].  The first application of this model on the 
spectrum of a HMXB (4U~0115$+$63) can be seen in [15].

\acknowledgments
AP acknowledges the Italian Space Agency financial support via contract 
I/008/07/0. This work has been partially supported by the Italian PRIN-INAF 2007 grant, 
"Bulk motion Comptonization models in X-ray Binaries: from phenomenology to physics", 
PI M. Cocchi.

\end{document}